# EUV Emission from Normal Galaxies

A. M. Read and T. J. Ponman
*School of Physics and Space Research, University of Birmingham, Edgbaston, BIRMINGHAM, B15 2TT*

**ABSTRACT**

Using data from the *ROSAT* Wide Field Camera all-sky survey, we have established upper limits to the extreme ultraviolet (EUV) flux from a sample of 30 bright, nearby, non-active spiral galaxies. These galaxies were chosen to be those most likely to be detected in the EUV on the basis of (i) low interstellar absorption within our own galaxy, (ii) brightness in other wavebands, (iii) high star formation activity, and (iv) proximity. The derived EUV upper limits are restrictive, and establish that the EUV flux escaping from galaxies does not contribute a major component of their bolometric luminosity, and in particular that it is very unlikely to be the sink for the energy injected into the interstellar medium by supernova explosions, as had been suggested following the failure to detect this power in the X-ray band.

## 1 INTRODUCTION

Galaxies with massive star-forming regions (starburst galaxies) are inferred to have very high supernova rates within their active regions. Rapidly expanding envelopes are ejected by these supernovae and interact with the surrounding interstellar medium (ISM), giving rise to hot gas. As this heated gas expands, it interacts with clouds which are compressed, heated by conduction, and are sometimes evaporated or disrupted. Eventually, these bubbles of hot gas surrounding the supernovae can rise far above the galactic plane and may produce a hot galactic corona before cooling and falling back to the plane or escaping the galaxy as a wind.

Recent *ROSAT* PSPC observations (Bregman & Pildis 1994, Walterbos et al. 1993), have now provided the first evidence for this hot gas around normal spiral galaxies. These galactic coronae though, seem to occur less frequently than had been expected. X-ray observations with *Einstein* for instance, failed to detect hot coronae, even in such nearby systems as the edge-on galaxies NGC 3628 and NGC 4244 (Bregman & Glassgold 1982) and the face-on spiral M101 (McCammon & Sanders 1984). Furthermore, for the few galactic halos that have been observed, the inferred soft X-ray luminosities of the halos account for only a very small fraction of the power injected into the galaxies by supernovae. In NGC 891 for instance, only ~1% of the supernova energy is being radiated in the halo (Bregman & Pildis 1994). The failure to detect large amounts of coronal or diffuse emission from these galaxies suggests that either the supernova energy is emerging in a spectral region unobservable both with *Einstein* and with the *ROSAT* PSPC, or it is spent in some radiationless process. The former requires emission from plasma at a temperature $T < 6 \times 10^5$ K (e.g. Walterbos et al. 1993), whilst the most likely non-radiative energy sink is kinetic energy loss in a galactic wind (Chevalier & Clegg 1985).

If the bulk of the supernova energy is radiated from a gas at a temperature below $6 \times 10^5$ K, it might be visible in the extreme ultra-violet (EUV). This region of the spectrum is subject to strong absorption by neutral and partially ionised gas. However the opacity of the interstellar medium is a strong function of wavelength (Cruddace et al. 1974), and extragalactic EUV emission at wavelengths <100 Å can penetrate the neutral hydrogen layer of our galaxy for sources at high galactic latitudes, as was demonstrated by the detection in the *ROSAT* Wide Field Camera (WFC) all-sky survey (Pye et al. 1995) of several active galaxies.

We have therefore undertaken a survey with the WFC of 30 nearby normal (i.e. non-active) galaxies, for which the chances of detecting EUV emission were judged to be highest, according to criteria specified in section 2. The analysis of the WFC data, described in section 3, resulted in no positive detections; however the upper limits derived for the EUV luminosities of these galaxies have interesting implications for the energy budget of galaxies, as is discussed in section 4.

## 2 OBSERVATIONS AND GALAXY SAMPLE

The *ROSAT* Wide Field Camera performed an all-sky survey in the EUV ($\approx 70 - 200$ eV) band during the period July 1990 – January 1991. The resulting catalogue of EUV sources, i.e the 2RE Source Catalogue (Pye et al. 1995), contains 479 sources, the majority of which are Galactic in origin. A number of extragalactic sources were detected in the routine survey processing, but all convincing identifications were with well known AGN, such as Mrk 421 and PKS 2155-304. We have reanalysed the survey data at the positions of a selected sample of non-active galaxies using improved source searching software. The sample studied included all those galaxies most likely to be detected in the EUV, on the basis of a variety of criteria discussed below.

Since EUV radiation is strongly attenuated by interstellar gas in our own Galaxy, the sample was restricted to spiral galaxies with $N_H < 3 \times 10^{20}$ cm$^{-2}$ (using the H I radio survey results of Stark et al. (1992)). As well as this $N_H$ condition being satisfied, the sample was selected on the basis of the following criteria. In all cases, distances are taken from Tully's Nearby Galaxy Catalog (1988). $H_0 = 75$ km s$^{-1}$ Mpc$^{-1}$ is assumed throughout.

### 2.1 Infrared brightness

One of the most interesting results from the *Infrared Astronomical Satellite* (IRAS) was the discovery of a class of luminous galaxies which emit the bulk of their energy in the far infrared (FIR). This infrared brightness is due primarily to dust heated by massive star-forming regions, and most of the galaxies in the sample have been selected on the basis of their IR flux. All non-active spiral galaxies with a 100 $\mu$m infrared flux greater than 64 Jy and a Galactic hydrogen column less than $3 \times 10^{20}$ cm$^{-2}$ are included in the sample (see Table 1). Non-active galaxies were distinguished from active galaxies using Condon & Broderick (1988), except in a few cases where subsequent discussion in the literature has changed their categorisation. The values of IR flux are taken from the IRAS bright galaxy sample (Soifer, Boehmer & Neugebauer 1989).

### 2.2 X-ray brightness

The *Einstein Observatory* gave us the capability to study galaxies in the soft X-ray band (0.2–3.5 keV) for the first time. Many of the previous IR-selected galaxies are also X-ray bright, but some members of the sample have been selected solely on the basis of their X-ray flux. These X-rays are thought to be due to X-ray binaries, supernova remnants, massive stars and shock-heated ISM material, all produced in large quantities during a starburst. All non-active spiral galaxies which met the $N_H$ criterion and have an X-ray flux greater than $45 \times 10^{-13}$ erg cm$^{-2}$ s$^{-1}$ are included in the sample (see Table 2). The values of X-ray flux (0.2–4.0 KeV) are taken from Fabbiano, Kim & Trinchieri (1992). Note that, unlike IRAS, *Einstein* did not carry out a complete sky survey, so the X-ray sample is not complete in any sense.

| Galaxy | 100 $\mu$m IR Flux (Jy) | Galactic Hydrogen Column ($10^{20}$ cm$^{-2}$) | Distance (Mpc) |
|---|---|---|---|
| N0253 | 1861.67 | 1.23 | 3.0 |
| N3079 | 103.40 | 0.76 | 20.4 |
| N3556 | 79.51 | 0.72 | 14.1 |
| N3627, M66, Arp317 | 137.44 | 1.87 | 6.6 |
| N3628, Arp317 | 105.96 | 1.77 | 7.7 |
| N3690, Arp299 | 122.45 | 0.83 | 41.4 |
| N4102 | 75.72 | 1.46 | 17.0 |
| N4254, M99 | 92.77 | 2.50 | 16.8 |
| N4303, M61 | 79.65 | 1.53 | 15.2 |
| N4321, M100 | 69.29 | 2.17 | 16.8 |
| N4414 | 69.11 | 1.35 | 9.7 |
| N4490, Arp269 | 85.94 | 1.63 | 7.8 |
| N4527 | 65.66 | 1.72 | 13.5 |
| N4631, Arp281 | 208.66 | 1.18 | 6.9 |
| N4666 | 82.88 | 1.60 | 14.1 |
| N4736, M94 | 121.52 | 1.28 | 4.3 |
| N4826, M64 | 77.66 | 2.35 | 4.1 |
| N5005 | 64.16 | 1.24 | 21.3 |
| N5055, M63 | 157.74 | 1.24 | 7.2 |
| N5194, M51, Arp85 | 280.40 | 1.25 | 7.7 |
| N5457, M101, Arp26 | 252.84 | 1.09 | 5.4 |

**Table 1.** The EUV survey: IR flux selected galaxies

| Galaxy | X-ray Flux (0.2–4.0keV) ($10^{-13}$ erg cm$^{-2}$s$^{-1}$) | Galactic Hydrogen Column ($10^{20}$ cm$^{-2}$) | Distance (Mpc) |
|---|---|---|---|
| N0253 | 51.80 | 1.23 | 3.0 |
| N4258 | 45.66 | 1.30 | 6.8 |
| N4579, M58 | 63.38 | 2.30 | 16.8 |
| N5033 | 77.37 | 0.97 | 18.7 |

**Table 2.** The EUV survey: X-ray flux selected galaxies

### 2.3 Supernova flux

The IR and X-ray selected galaxies above are all expected to have large values of 'supernova flux' (i.e $f_{SN} = L_{SN}/4\pi d^2$, where $L_{SN}$ is the estimated supernova luminosity (section 3.4), and $d$ is the distance), since the IR and/or X-ray excess they show is linked to the amount of supernova activity within the galaxy. As one of the major aims of the present survey was to study the relationship between EUV luminosity and the energy released into galaxies by supernova explosions, a number of galaxies which did not qualify as X-ray or IR bright were added to the sample. These galaxies were chosen on the grounds that should have large values of supernova flux because of their proximity and favourable hydrogen columns. These galaxies are listed in Table 3.

| Galaxy | Galactic Hydrogen Column ($10^{20}$ cm$^{-2}$) | Distance (Mpc) |
|---|---|---|
| N0055 | 1.34 | 1.3 |
| N0247 | 1.33 | 2.1 |
| N0300 | 2.97 | 1.2 |
| N1291 | 1.79 | 8.6 |
| N3184 | 1.02 | 8.7 |
| N7793 | 1.03 | 2.8 |

**Table 3.** The EUV survey: Further galaxies with high 'supernova flux'.

## 3 DATA REDUCTION AND ANALYSIS

The analysis of all galaxies within the present survey has been performed in a uniform manner using the Starlink *AS-TERIX* X-ray analysis system, as described below.

### 3.1 WFC data reduction

Two filters were used during the WFC all-sky survey, an S1 filter with a mean energy of 124 eV and a bandpass of 90−185 eV (at 10% of the peak), and an S2 filter with a mean energy of 90 eV and a bandpass of 62−111 eV. Interstellar absorption makes detection of extragalactic objects extremely unlikely in the S2 filter, hence only data from the shorter wavelength S1 filter were used in the present study.

The photons from a 2° × 2° area were binned into a raw image and then background subtracted. All of the fields obtained during the all-sky survey had been processed using a standard analysis system. This standard WFC survey analysis however, implicitly assumes that sources are unresolved and since the 50% enclosed energy radius of the survey Point Spread Function (PSF) is only 2′, it can be seen from the galaxy sizes in Table 4 that this is not a good assumption. Hence the sensitivity of the search was tuned to the expected galaxy diameters by blurring the nominal instrument PSF with a Gaussian of FWHM equal to the optical diameter of each system.

The point source search program PSS (Allan, Ponman & Jeffries 1995), was used to search for emission from these galaxies. This uses a likelihood method to search for enhancements above a specified background, arising in the present case from the internal detector background, high-energy charged particles associated with the Earth's radiation belts and cosmic rays, low energy electrons, geocoronal and atmospheric EUV and UV line emission, and diffuse EUV sky background. No significant emission was detected from any of the 30 systems and so 90% confidence upper limits to the WFC count rates within the optical radius of each galaxy were derived at the positions given in the Second Reference Catalog of Bright Galaxies (hereafter, RCBG) (de Vaucouleurs, de Vaucouleurs & Corwin 1976). These limits can be found in Table 4.

To convert the count rate upper limits into upper limits to the flux and hence to luminosity, a spectral model has to be assumed. A hot optically thin plasma is the most likely source of any diffuse EUV emission from an external galaxy and so Raymond & Smith (1977) hot plasma models (with solar metallicity) were assumed. Two models, with different temperatures were used: one ($5 \times 10^5$ K) representing the highest approximate temperature invisible to both *Einstein* and the *ROSAT* PSPC, and the other ($2 \times 10^5$ K) at the low temperature limit of modelled galactic coronae (see section 4.2).

### 3.2 Interstellar absorption

Three different absorption models have also been used, as follows:

- In model A1, only absorption within in our own galaxy is considered. This is incorporated when calculating the observed count rate, but removed when calculating the flux. This case corresponds to emission from a corona lying above the absorbing layers of the host galaxy, and the calculated luminosity is that leaving the target galaxy.
- Model A2 includes absorption from both our own and the host galaxy. This host absorbing material is assumed to 'sandwich' the layer from which the EUV radiation is emitted. The calculation of the amount of absorbing material within the host systems is described below. The luminosity tabulated is that emitted *within* the host galaxy, but most of this is actually absorbed internally. Hence the resulting upper limits are very high.
- Model A3 includes the same galactic and host columns as A2, but in this case the luminosity tabulated is that which actually escapes from the galaxy.

Models A1 and A3 may, at first sight, appear identical. Within model A3, however, the flux emitted from the disc of the host galaxy is absorbed by the host hydrogen column and much of the low energy flux is lost. Consequently, little of the emergent flux is absorbed by our own Galaxy as most of the soft emission has already been absorbed. There is a very significant *spectral* difference between the emergent flux in models A1 and A3.

For models A2 and A3, an estimate of the column within the target galaxies is needed. We have used results from the 18 cm OH and 21 cm H I survey of 88 luminous galaxies by Martin et al. (1991). This demonstrated that, although total mass and total blue luminosity increase with increasing FIR luminosity, high-$L_{FIR}$ galaxies are deficient in their relative H I content. Their mean H I projected densities, for example, are significantly smaller than in classical galaxies of the Hubble sequence. This deficiency appears to be correlated with star formation activity.

We have fitted regression lines to their data for the mean H I surface brightness as a function of FIR luminosity for samples of both early and late-type spirals. Although the plots show a large degree of scatter, for spirals of type T=0−3 (early-type spirals), the mean relationship is

$$N_H = 10^{[22.0 - 0.10 \log L_{FIR}/L_\odot]} \text{ cm}^{-2}, \quad (1)$$

whereas for late-type spirals (T ≥ 4)

$$N_H = 10^{[21.3 - 0.01 \log L_{FIR}/L_\odot]} \text{ cm}^{-2}. \quad (2)$$

As can be seen from these equations, the mean H I surface density is reduced as the FIR luminosity is increased, though this effect is not so prominent in the later spirals.

| Galaxy | Sample | | Properties | | | WFC upper limit | Columns | |
|---|---|---|---|---|---|---|---|---|
| | (I,X or O) | (*) | Distance (Mpc) | Axial ratio (D/d) | Major diam. (arcmin.) | (counts ks$^{-1}$) | Host | Galactic |
| | | | | | | | ($10^{20}$ cm$^{-2}$) | |
| N0055 | O |   | 1.3  | 5.01 | 32.4 | 41.45 | 32.62 | 1.33 |
| N0247 | O |   | 2.1  | 2.69 | 20.0 | 10.09 | 17.73 | 1.33 |
| N0253 | I X | * | 3.0  | 3.39 | 25.1 | 24.40 | 20.85 | 1.23 |
| N0300 | O |   | 1.2  | 1.35 | 20.0 | 80.86 | 8.79  | 2.97 |
| N1291 | O |   | 8.6  | 1.15 | 10.5 | 14.56 | 7.11  | 1.79 |
| N3079 | I | * | 20.4 | 4.47 | 7.6  | 8.42  | 27.27 | 0.76 |
| N3184 | O |   | 8.7  | 1.02 | 6.9  | 2.94  | 6.64  | 1.02 |
| N3556 | I | * | 14.1 | 3.31 | 8.3  | 8.11  | 20.52 | 0.72 |
| N3627 | I | * | 6.6  | 1.99 | 8.7  | 5.79  | 9.15  | 1.87 |
| N3628 | I |   | 7.7  | 4.07 | 14.8 | 11.20 | 19.25 | 1.77 |
| N3690 | I | * | 41.4 | 1.26 | 2.4  | 1.55  | 7.40  | 0.83 |
| N4102 | I | * | 17.0 | 1.62 | 3.2  | 12.83 | 6.95  | 1.46 |
| N4254 | I | * | 16.8 | 1.12 | 5.4  | 8.68  | 6.91  | 2.50 |
| N4258 | X |   | 6.8  | 2.29 | 18.2 | 5.29  | 14.91 | 1.30 |
| N4303 | I | * | 15.2 | 1.07 | 6.0  | 8.42  | 6.51  | 1.53 |
| N4321 | I |   | 16.8 | 1.12 | 6.9  | 3.60  | 6.95  | 2.17 |
| N4414 | I | * | 9.7  | 1.66 | 3.6  | 7.92  | 10.34 | 1.34 |
| N4490 | I | * | 7.8  | 1.91 | 5.9  | 13.16 | 12.03 | 1.63 |
| N4527 | I | * | 13.5 | 2.75 | 6.3  | 5.24  | 16.61 | 1.72 |
| N4579 | X |   | 16.8 | 1.23 | 5.4  | 6.42  | 7.60  | 2.30 |
| N4631 | I | * | 6.9  | 4.57 | 15.1 | 7.56  | 28.52 | 1.18 |
| N4666 | I | * | 14.1 | 2.95 | 4.5  | 8.45  | 17.82 | 1.60 |
| N4736 | I |   | 4.3  | 1.20 | 11.0 | 9.38  | 6.60  | 1.28 |
| N4826 | I |   | 4.1  | 1.74 | 9.3  | 4.63  | 10.16 | 2.35 |
| N5005 | I | * | 21.3 | 2.00 | 5.4  | 5.18  | 12.28 | 1.24 |
| N5033 | X |   | 18.7 | 1.86 | 10.5 | 7.93  | 12.11 | 0.97 |
| N5055 | I |   | 7.2  | 1.62 | 12.3 | 4.92  | 10.16 | 1.24 |
| N5194 | I | * | 7.7  | 1.41 | 11.0 | 9.14  | 8.69  | 1.25 |
| N5457 | I |   | 5.4  | 1.02 | 26.9 | 8.79  | 6.31  | 1.09 |
| N7793 | O |   | 2.8  | 1.38 | 9.1  | 22.62 | 8.98  | 1.03 |

**Table 4.** Some key properties of the sample galaxies, and the WFC count rate upper limits (90%). The 'sample' notation is: I = infrared bright, X = X-ray bright, O = additional galaxy added on the basis of high predicted supernova flux at the Earth. Starburst galaxies (here, $L_{FIR} > 0.38 L_B$) are marked with an asterisk.

In order to arrive at a value for the H I column density to the central plane of each inclined galactic disc (i.e through one half of the 'sandwich'), the host columns calculated from the above equations were divided by two and multiplied by the axial ratio (from RCBG) of each galaxy (the values of Martin et al. are inclination corrected). These values are tabulated as the host columns in Table 4.

### 3.3 Multiwavelength luminosities

In addition to the six EUV upper limits (bolometric luminosities) given in Table 5, a number of other luminosities are tabulated. These all use distances from Tully (1988), which are based on $H_0 = 75$ km s$^{-1}$ Mpc$^{-1}$, and assume that the Galaxy is retarded by 300 km s$^{-1}$ from universal expansion by the mass of the Virgo cluster.

Optical (B) luminosities for all the galaxies in the sample are taken from Tully (1988). FIR luminosities were calculated from IRAS 60 and 100 $\mu$m fluxes using the expression

$$L_{FIR} = 3.65 \times 10^5 \left[ 2.58 S_{60\mu m} + S_{100\mu m} \right] D^2 L_\odot,$$

(e.g. Devereux & Eales 1989) where $D$ is the distance in Mpc and $S_{60\mu m}$ and $S_{100\mu m}$, the IRAS 60 and 100 $\mu$m fluxes (in Janskys), are taken from Soifer et al. (1989), and supplemented for the nearby, low-IRAS flux cases, from Rice et al. (1988). The values of (0.2−4.0 keV) X-ray luminosity are taken from Fabbiano et al. (1992), and scaled to the Tully (1988) distances.

### 3.4 Supernova luminosities

'Supernova luminosities' for the 30 galaxies in the sample, i.e the *power* injected into each galaxy by supernovae, were calculated using the inclination corrected supernova rates (types I and II combined) for each morphological type given in van den Bergh & Tammann (1991). The suggestion of these authors that the rates should be simply scaled with the blue luminosity is unsatisfactory for starburst galaxies, in which optical emission from the active star-forming regions is subject to substantial obscuration by dust. As a result, $L_B$ underestimates the star formation rate, and hence the supernova rate.

We have therefore distinguished between normal and starburst galaxies (here defined as those with $L_{FIR}/L_B > 0.38$). Supernova luminosities for the former were obtained by simply scaling with $L_B$, using the definition of super-nova rate given in van den Bergh & Tammann (1991) and assuming an energy release of $10^{51}$ ergs per supernova,

$$L_{SN} \text{ (non-starburst)} = 0.00465 L_B S \text{ erg s}^{-1}, \quad (3)$$

where $S$, the number of supernovae per century per $10^{10} L_\odot(B)$, is the appropriate value from van den Bergh & Tammann (1991).

| Galaxy | Log Luminosity (erg s$^{-1}$) | | | | | | | | | |
|---|---|---|---|---|---|---|---|---|---|---|
| | $L_B$ | $L_{FIR}$ | $L_X$ | $L_{SN}$ | Bolometric $L_{EUV}$ upper limits | | | | | |
| | | | | | T= $2 \times 10^5$ K | | | T= $5 \times 10^5$ K | | |
| | | | | | A1 | A2 | A3 | A1 | A2 | A3 |
| N0055 | 43.02 | 41.94 | - | 41.44 | 43.37 | 52.75 | 41.44 | 40.96 | 47.89 | 42.33 |
| N0247 | 42.96 | 41.48 | 38.86 | 41.34 | 43.18 | 50.28 | 40.74 | 40.76 | 46.02 | 41.43 |
| N0253 | 43.78 | 43.74 | 39.75 | 42.62 | 43.74 | 51.49 | 41.53 | 41.36 | 47.01 | 42.18 |
| N0300 | 42.52 | 41.43 | - | 40.90 | 45.19 | 49.24 | 41.20 | 42.40 | 45.38 | 41.69 |
| N1291 | 43.88 | 42.18 | - | 41.39 | 45.09 | 49.35 | 41.71 | 42.53 | 45.77 | 42.31 |
| N3079 | 44.18 | 44.14 | 40.47 | 43.02 | 44.26 | 53.62 | 42.88 | 42.09 | 49.01 | 43.74 |
| N3184 | 43.51 | 42.75 | - | 41.89 | 43.45 | 48.17 | 40.65 | 41.16 | 44.78 | 41.38 |
| N3556 | 44.08 | 43.67 | - | 42.49 | 43.86 | 52.22 | 42.30 | 41.71 | 47.88 | 43.07 |
| N3627 | 43.64 | 43.29 | - | 41.73 | 44.54 | 49.38 | 41.26 | 41.96 | 45.57 | 41.84 |
| N3628 | 43.76 | 43.31 | 40.07 | 41.75 | 44.86 | 51.79 | 42.04 | 42.31 | 47.49 | 42.78 |
| N3690 | 44.17 | 45.03 | 41.00 | 43.97 | 44.25 | 49.70 | 42.10 | 42.04 | 46.19 | 42.56 |
| N4102 | 43.38 | 43.93 | - | 42.45 | 45.25 | 49.70 | 42.10 | 42.79 | 46.19 | 42.75 |
| N4254 | 44.11 | 43.87 | 40.19 | 42.72 | 46.12 | 49.89 | 42.30 | 43.41 | 46.24 | 42.80 |
| N4258 | 44.01 | 42.95 | 40.40 | 42.39 | 43.88 | 50.51 | 41.38 | 41.47 | 46.44 | 42.07 |
| N4303 | 44.06 | 43.76 | 40.42 | 42.61 | 45.06 | 49.27 | 41.79 | 42.58 | 45.82 | 42.44 |
| N4321 | 44.22 | 43.74 | 40.45 | 42.60 | 45.44 | 49.40 | 41.80 | 42.79 | 45.79 | 42.36 |
| N4414 | 43.42 | 43.29 | - | 42.17 | 44.41 | 50.03 | 41.67 | 41.99 | 46.05 | 42.29 |
| N4490 | 43.59 | 43.26 | - | 42.10 | 44.79 | 50.51 | 41.85 | 42.27 | 46.46 | 42.46 |
| N4527 | 43.66 | 43.59 | - | 42.47 | 44.96 | 51.49 | 42.11 | 42.42 | 47.24 | 42.76 |
| N4579 | 44.04 | 43.16 | 41.33 | 42.03 | 45.81 | 49.91 | 42.15 | 43.14 | 46.21 | 42.68 |
| N4631 | 43.82 | 43.46 | 39.90 | 42.29 | 43.89 | 52.87 | 41.99 | 41.53 | 48.19 | 42.86 |
| N4666 | 43.68 | 43.71 | - | 42.59 | 45.07 | 51.92 | 42.37 | 42.57 | 47.72 | 43.12 |
| N4736 | 43.52 | 42.90 | 39.66 | 41.51 | 43.70 | 48.16 | 40.65 | 41.30 | 44.73 | 41.34 |
| N4826 | 43.43 | 42.59 | 39.20 | 41.42 | 44.49 | 49.24 | 40.92 | 41.81 | 45.30 | 41.47 |
| N5005 | 44.28 | 43.90 | - | 42.73 | 44.78 | 50.95 | 42.24 | 42.40 | 47.04 | 42.89 |
| N5033 | 44.18 | 43.67 | 41.51 | 42.56 | 44.48 | 50.93 | 42.25 | 42.21 | 46.92 | 42.91 |
| N5055 | 43.83 | 43.06 | - | 42.21 | 43.82 | 49.49 | 41.17 | 41.44 | 45.65 | 41.80 |
| N5194 | 44.07 | 43.66 | 40.39 | 42.48 | 44.16 | 49.40 | 41.38 | 41.77 | 45.69 | 42.03 |
| N5457 | 43.95 | 43.30 | 39.83 | 42.33 | 43.62 | 48.12 | 40.69 | 41.29 | 44.77 | 41.42 |
| N7793 | 42.85 | 42.07 | 38.81 | 41.27 | 43.37 | 48.94 | 40.86 | 41.08 | 45.27 | 41.52 |

**Table 5.** Multiwavelength luminosities derived as described in the text. Bolometric EUV luminosities are 90% confidence upper limits for sources similar to the optical size of the galaxy.

For the starburst galaxies, an estimate of the component of $L_{FIR}$ associated with star formation (as opposed to that associated with the interstellar radiation field) needs to be used to scale the supernova rate. This estimate was derived using the method of Devereux & Eales (1989), where the luminosity imparted to dust grains by the interstellar radiation field is assumed to be a constant fraction, $f$, of the blue luminosity $L_B$. Hence, the *star-formation* luminosity, $L_{SF}$, can be expressed as

$$L_{SF} = L_{FIR} - f L_B. \quad (4)$$

This yields an expression for the supernova luminosities for the starburst galaxies, again assuming $10^{51}$ ergs per supernova,

$$L_{SN} \text{ (starburst)} = 0.0153 \left[ L_{FIR} - 0.076 L_B \right] S \text{ erg s}^{-1}, \quad (5)$$

where $S$ is again the appropriate value from van den Bergh & Tammann (1991). For galaxies of an intermediate activity (i.e $L_{FIR} \approx 0.38 L_B$), equations 3 and 5 give approximately the same value of $L_{SN}$. It should be noted that $S$ is greater for later-type spirals than for earlier-types. This has been taken into account when calculating $L_{SN}$ for each galaxy.

The resultant supernova luminosities for the sample, together with other multiwavelength information including the six bolometric $L_{EUV}$ upper limits are shown in Table 5.

## 4 RESULTS AND DISCUSSION

### 4.1 The broadband spectra of galaxies

For the cases in which we are considering the actual amount of EUV emission *escaping* from the host galaxies (i.e models A1 and A3), Table 5 shows that for the high temperature case, these galaxies must generally emit less radiation in the EUV than they do in either the infrared or the optical bands. In the majority of cases, whatever EUV luminosity is emitted is more than an order of magnitude below the infrared and optical luminosities. Even in the low temperature case, this is seen to be true for the A3 model, though not for A1 (the model corresponding to a hot corona). However, as shall be discussed below, both recent observations and current theoretical models suggest that the bulk of the coronal gas cannot radiate at temperatures as low as $2 \times 10^5$ K. Hence these WFC results establish that, for nearby non-active galaxies (both normal and starburst), the EUV emission is very unlikely to be a dominant component of the emitted radiation.

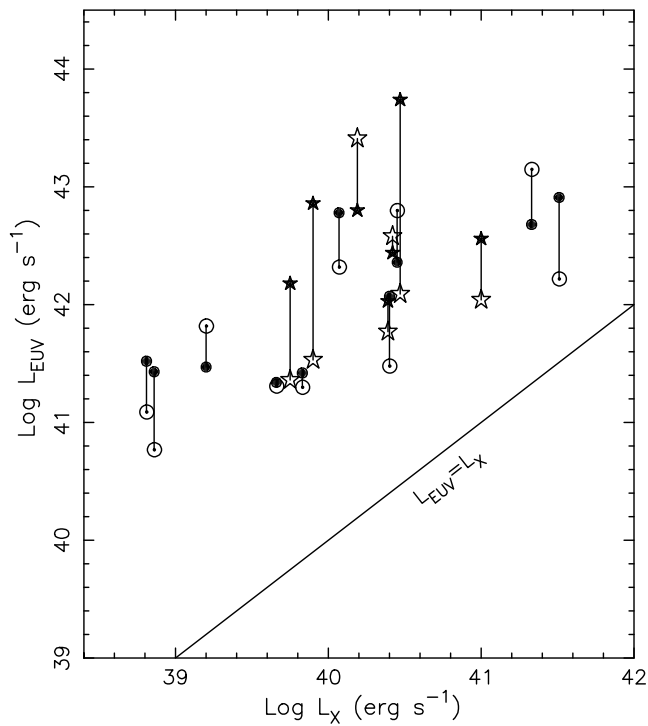

**Figure 1.** Comparison between bolometric EUV upper limits for emission escaping from galaxies (assuming T = $5 \times 10^5$ K and using absorption models A1 and A3) and their (0.2–4.0 keV) X-ray luminosities. Circles denote normal galaxies and stars denote starbursts. The filled symbols represent the A3 results, the open symbols represent the A1 results. The two EUV upper limits for each galaxy are joined by a line.

Fig. 1 shows the comparison between the limits on the escaping EUV radiation (models A1 and A3, with $T = 5 \times 10^5$ K) and the X-ray luminosity. The EUV results are bolometric, whilst the X-ray luminosities are in the range 0.2–4.0 keV. Whereas the rather hard galactic X-ray spectra used in the past (e.g. David, Jones & Forman 1992) result in bolometric values a factor 2–5 higher, recent *ROSAT* low temperature spectra give bolometric correction factors of order unity. The A1 and A3 models represent the two extreme scenarios for the origin of escaping EUV radiation, i.e emission from the galactic halo and emission from the very central parts of the disc. What can be concluded from this figure, remembering that these $L_{EUV}$ values are upper limits, is that, in the most tightly constrained cases, the escaping EUV luminosity is at most an order of magnitude greater than the X-ray luminosity.

In Fig. 2 we show the broad band spectra for two galaxies for which the upper limits are particularly tight. NGC 5457 is a fairly normal galaxy viewed face-on, whereas NGC 3690 is a well known starburst galaxy. In both cases it is evident that the energy budget is dominated by the optical and the FIR. Note in particular that the EUV luminosity falls well short of the calculated supernova luminosity, $L_{SN}$.

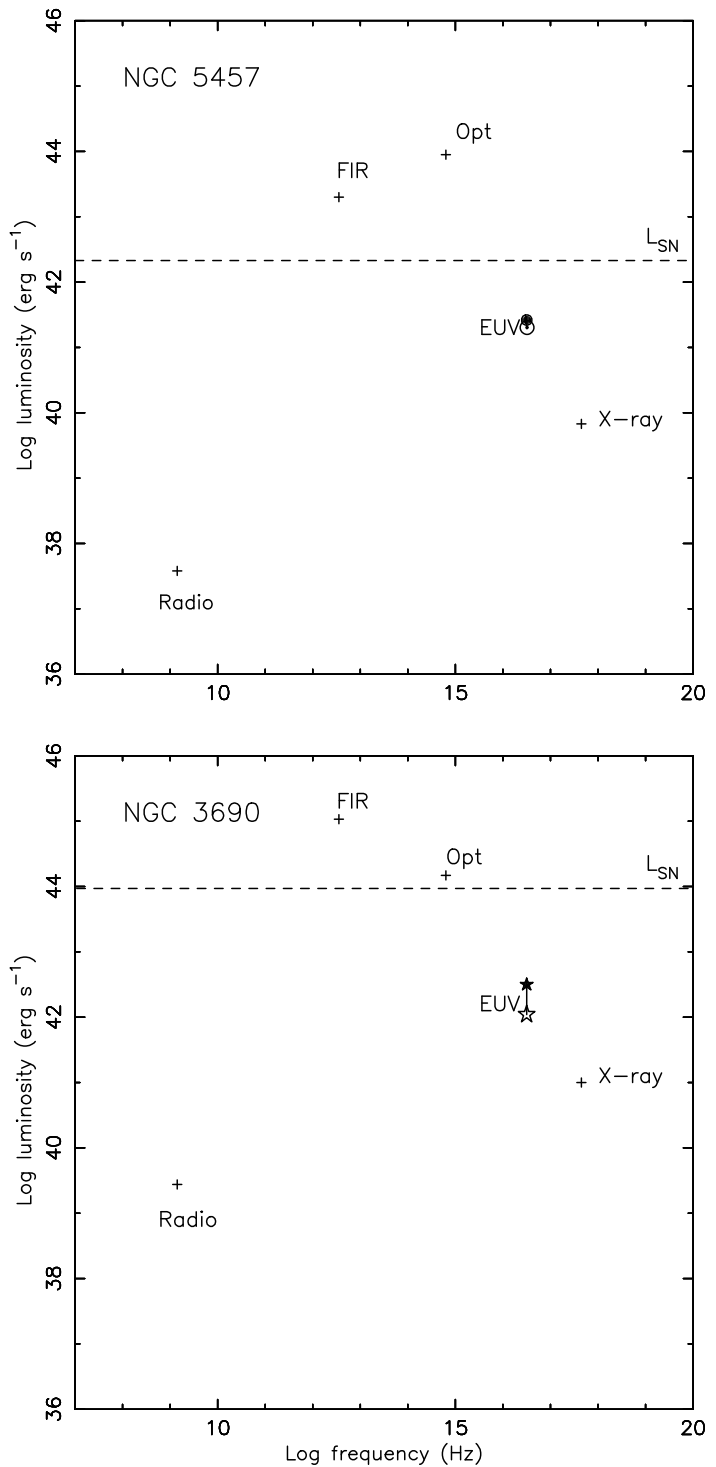

**Figure 2.** Multiwavelength spectra of (top) the normal galaxy NGC 5457, and (bottom) the starburst galaxy NGC 3690. Radio luminosities are $\nu L_\nu$, where $L_\nu$ is the 1.49 GHz spectral luminosity from Condon et al. (1990), scaled to the Tully distances. The EUV points are upper limits for bolometric flux escaping from the galaxy, i.e models A1 (open symbols) and A3 (filled symbols) for T = $5 \times 10^5$ K. A line is drawn at the level of $L_{SN}$ (calculated as in section 3.4) as it is not known in what region of the spectrum this energy appears.

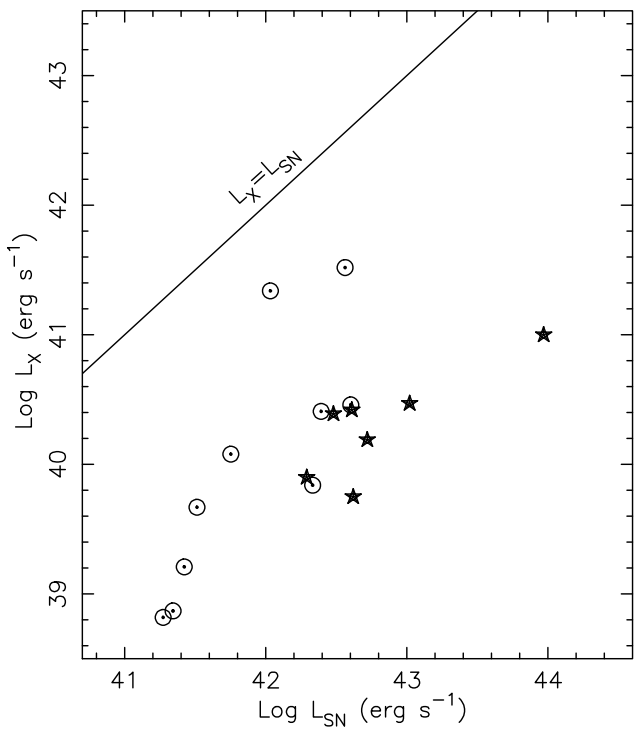

**Figure 3.** Comparison between X-ray and supernova luminosities for all members of the sample for which X-ray luminosities (Fabbiano et al. 1992) are available. Circles denote normal galaxies and stars denote starbursts.

### 4.2 Coronal emission

The existence, within the planes of spirals, of *superbubbles*, containing hot, low-density gas is now well established observationally. It is, however, not yet clear how frequently such bubbles are able to break out of the plane and inject hot gas into a galactic corona. Such a corona would act as a sink for the substantial supernova luminosity apparent both in Table 5 and in Fig. 2, and would also account for the existence of high ionization ions (such as Nv) observed at high altitudes, for high velocity clouds above the plane, and for at least part of the soft X-ray background. These features have motivated a variety of *galactic fountain* models (Shapiro & Field 1976), in which hot gas wells up into the corona (either freely or through *chimneys* (Norman & Ikeuchi 1989) established where superbubbles break out of the plane). It then cools and falls back into the disc.

As mentioned previously, X-ray observations of certain nearby galaxies with both *Einstein* (Bregman & Glassgold 1982; McCammon & Sanders 1984) and *ROSAT* (Bregman & Pildis 1994; Walterbos et al. 1993) have already demonstrated that the X-ray luminosity is *not* sufficient to account for the supernova energy. It can be seen from Fig. 3 that this also holds for all the galaxies in our sample for which X-ray luminosities are available. This rules out emission of the supernova luminosity from any corona with $T > 6 \times 10^5$ K.

It can be seen from Table 5 that, in many cases, the EUV upper limits rule out the possibility that the bulk of the available supernova energy can be radiated away by a $5 \times 10^5$ K corona. This is shown graphically in Fig. 4, where the cumulative distribution of $L_{EUV}/L_{SN}$ ratios for the galaxy sample is shown. In the $5 \times 10^5$ K case, the EUV limit is less than the supernova luminosity for 20 of the 30 galaxies (and for 11 of the 15 starbursts). In the $2 \times 10^5$ K case, however, the $L_{EUV}$ limit is greater than $L_{SN}$ for all of the systems, allowing the possibility that the bulk of the supernova energy is radiated away by a cooler, $2 \times 10^5$ K corona.

Current theoretical modelling and recent observations of our own Galaxy however, suggest that the bulk of the energy in galactic halos should be radiated at temperatures higher than this. A variety of hydrodynamical computational studies (e.g. Bloemen 1987; Houck & Bregman 1990; Shull & Slavin 1994) predicts coronae for which the emission measure lies largely or completely in gas at temperatures of $3 \times 10^5 - 10^6$ K.

There is also *observational* evidence that the temperature of the Galactic halo gas cannot be as low as $2 \times 10^5$ K. The observed flux in the Galactic halo CIV lines, for instance, agrees quantitatively with that expected from a halo gas cooling radiatively through a temperature of a few $\times 10^5$ K. A cooler gas, at a temperature $< 1 - 2 \times 10^5$ K, could not readily fit the data, since electrons at this temperature would have a mean energy far short of the 8.0 eV needed to excite $C^{+3}$ ions (Martin & Bowyer 1990). Edgar & Chevalier (1986) have also shown that Galactic halo absorption lines like Nv are most likely to be formed in a gas cooling from temperatures higher than $2 \times 10^5$ K.

Modelling of a $3 \times 10^5$ K corona, perhaps the lowest realistic temperature one could consider, results in A1 (coronal emission) $L_{EUV}$ limits less than the supernova luminosity for two of the galaxies. The results are to some extent sensitive to assumed metallicity, in the sense that low metallicity gas results in lower limits to $L_{EUV}$. It is also worth noting that smaller galaxies (73% of our sample galaxies are smaller than the Milky Way) tend to have lower metallicity, and hence supernova remnants should break out of the plane more easily due to both the shallower potential well and to less efficient cooling. Hence any EUV emitting corona is even less likely to be cool in such systems.

### 4.3 The supernova energy sink

If the supernova energy does not escape in the EUV, then what happens to it?

One possible sink for the energy injected by supernovae, is in the ejection of High Velocity Clouds (HVCs) of neutral hydrogen, with masses up to $10^8 M_\odot$, high into the galactic halo. A recent HI survey of several face-on disc galaxies however, has established that, although evidence for large amounts (up to a few $\times 10^9 M_\odot$) of high velocity material is seen within these systems, the kinetic energy involved is a small fraction of the injected supernova energy (less than 1% in the majority of cases) (Schulman, Bregman & Roberts 1994).

Another possibility is that the supernovae drive a hot galactic wind, losing energy in the form of kinetic and thermal energy of ejected material. This seems most unlikely for non-starburst galaxies, given the large number of correlated supernovae believed to be required to break out of the gas layer in the plane (Koo & McKee 1992), and the lack of any observational evidence for winds in such systems. However observations of extended soft X-ray emission and optical emission line nebulae (Heckman, Lehnert & Armus 1993) indicate that vigorous winds are present in many starburst galaxies, and should carry off a substantial fraction of the energy released by supernovae in the central starburst.

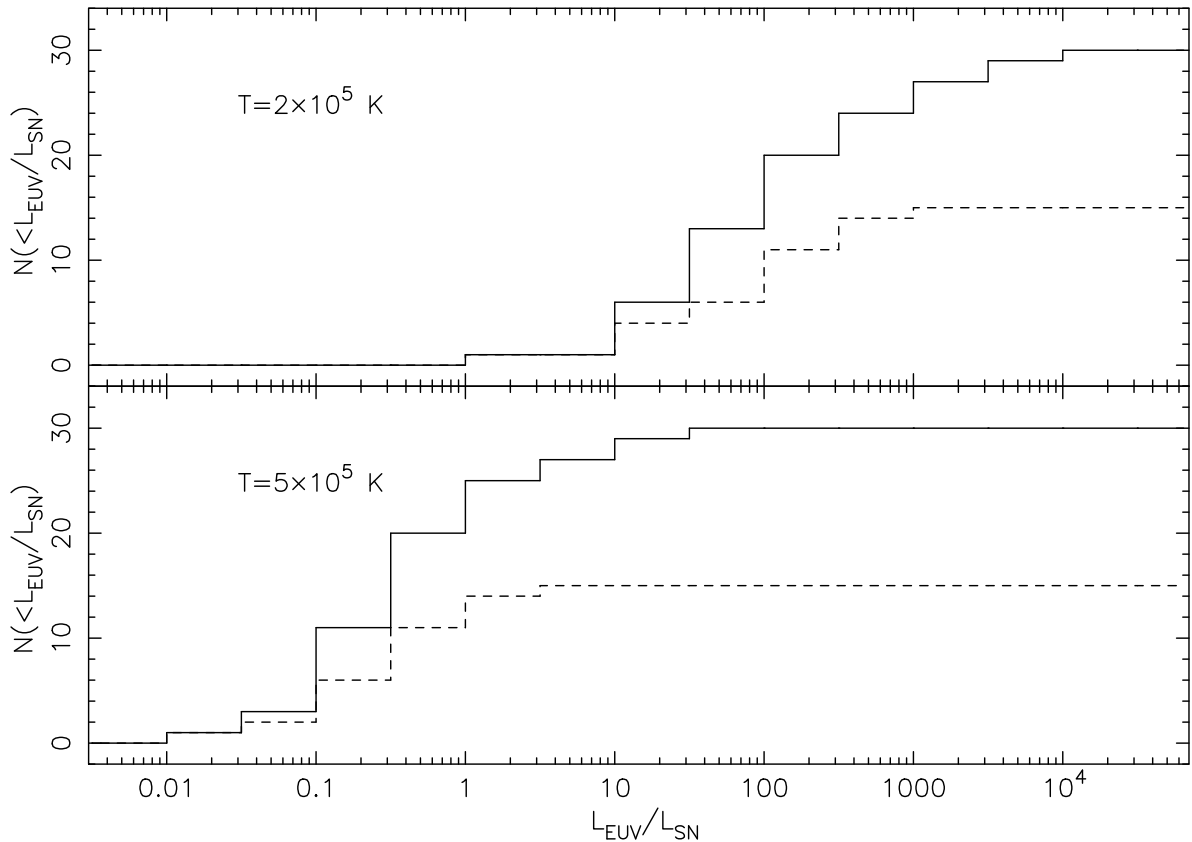

**Figure 4.** Histograms showing the cumulative distribution of $L_{EUV}/L_{SN}$ upper limits for the galaxy sample. The results for absorption model A1 (i.e coronal emission) are shown. Limits are given for two different plasma temperatures, and the distribution of starbursts in the sample is shown dashed. It can be seen that in the high temperature case, the $L_{EUV}$ upper limit is less than the power injected into the system by supernovae for 20 of the galaxies (of which 11 are starbursts).

If, as seems probable for most galaxies, the energy loss in galactic winds is unable to account for the bulk of the supernova luminosity, then the most likely possibility is that it is initially lost in the EUV by supernova remnants in the radiative phase, but is absorbed locally and degraded to lower energies, where it is hidden in the large UV-optical-IR flux, and will also add to the FIR emission through irradiation of dust. Fig. 5 suggests that if all the supernova flux were emitted in the EUV close to the disc plane, it could easily be attenuated within the absorbing gas layer of the host galaxy to a point where it is consistent with our non-detections.

However, the simple assumption implicit in our A2 model, of absorption in the host galaxy by a single disc-averaged $N_H$, may overestimate the opacity to EUV emission. Most galaxies, especially starburst and interacting galaxies, show a deficiency of H<small>I</small> in their inner regions, where the star formation peaks. Furthermore, the H<small>I</small> distribution in the discs of nearby galaxies is punctuated by 'H<small>I</small> holes', which can cover 10-20% of the disc, even in normal galaxies (Heiles 1990). If, as is commonly supposed, these holes represent regions from which H<small>I</small> has been cleared by supernova bubbles, then they should coincide with the sites of much of the supernova-related EUV and X-ray emission. Whilst we have to bear in mind that the main source of opacity in the EUV band is actually He, rather than H<small>I</small>, it is interesting that a face-on galaxy such as NGC 5457, where many H<small>I</small> holes are seen (e.g. van der Hulst & Sancisi 1988; Kamphuis, Sancisi & van der Hulst 1991), should have escaping X-ray and EUV luminosities which are both well below $L_{SN}$ (Fig. 2), and more sensitive studies in the EUV may be able to set interesting constraints on the evolution of superbubbles in such systems.

## 5 CONCLUSIONS

Restrictive upper limits on the EUV emission from a sample of normal galaxies, chosen to be those most likely to be detected in the EUV, have been derived using data from the

<="">*EUV Emission from Normal Galaxies* 9</>

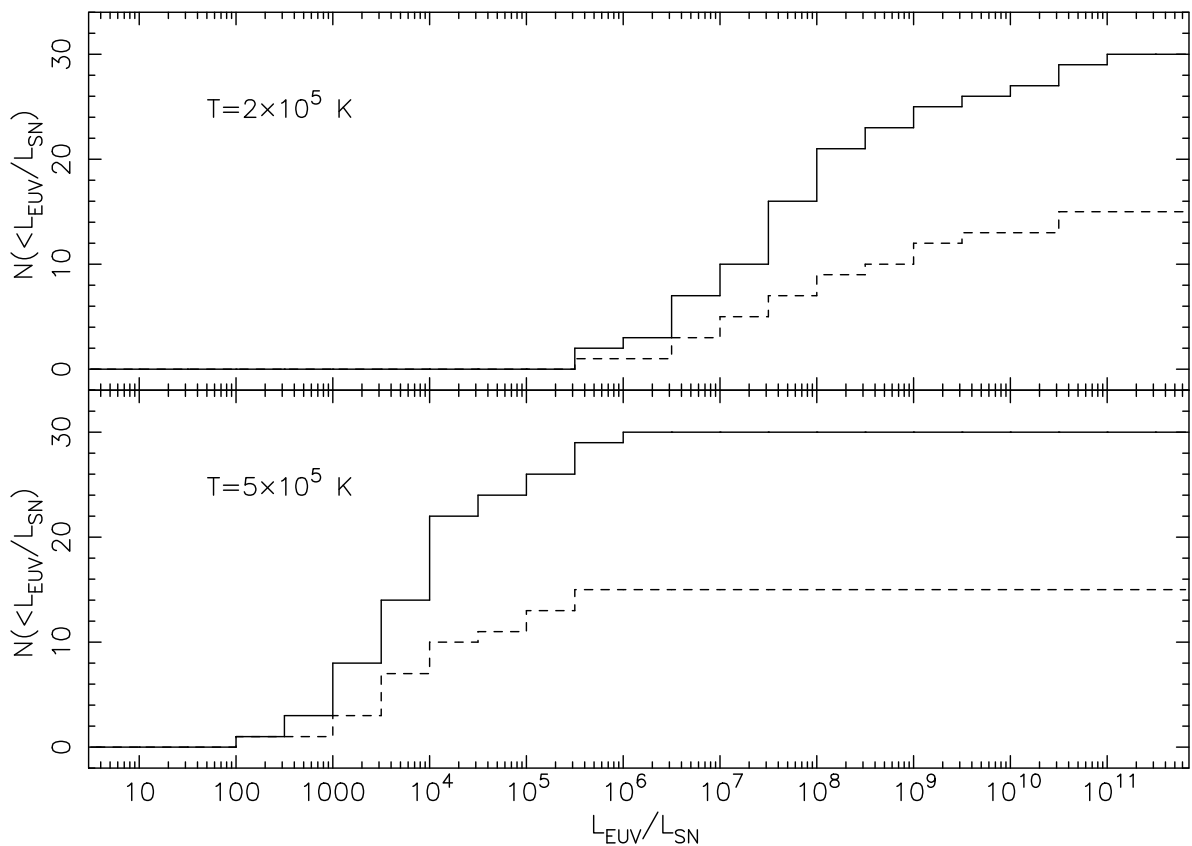

**Figure 5.** As Fig. 4, except that the EUV limit is for emission *within* the host galaxy (model A2). Limits are given for two different plasma temperatures, and the distribution of starbursts in the sample is shown dashed. $L_{EUV} = L_{SN}$ is allowed in all cases.

WFC all-sky survey. These limits establish that the EUV emission from normal galaxies cannot be a major component of their bolometric luminosity unless it is radiated by gas with a characteristic temperature $\lesssim 2 \times 10^5$ K. Since it is believed that the bulk of the gas in galactic coronae should be hotter than this, one important implication is that the energy liberated by supernovae cannot be radiated in the EUV from a hot corona, as was suggested following the failure to detect this power in the X-ray. It seems most likely that it is degraded and escapes in the UV, optical and infrared, though galactic winds may also constitute a significant energy loss for starburst galaxies.

## ACKNOWLEDGEMENTS

AMR acknowledges the receipt of a SERC/PPARC studentship. We also thank the referee for useful comments. Data reduction and analysis was performed on the Starlink node at Birmingham.